\documentclass[11pt]{article}
\usepackage{moriond}

\usepackage{amsmath}
\usepackage{amssymb}

\def\be{\begin{equation}}
\def\ee{\end{equation}}
\def\bea{\begin{eqnarray}}
\def\eea{\end{eqnarray}}

\newcommand{\Photo}{}

\begin{document}
\vspace*{4cm}

\title{New Physics Interpretations of the $B \to K^* \mu^+\mu^-$ Anomaly}
\author{ Wolfgang Altmannshofer }
\address{Perimeter Institute for Theoretical Physics, \\ 31 Caroline Street North, Waterloo, Ontario N2L 2Y5, Canada}

\maketitle\abstracts{This talk discusses possible new physics interpretations of recent experimental results on the $B\to K^* \mu^+\mu^-$ decay that show a discrepancy with the Standard Model predictions. 
A model independent analysis that takes into account all the relevant observables in $B\to K^* \mu^+\mu^-$ and in related $b \to s$ transitions allows to identify a consistent new physics explanation of the discrepancy.
An explicit realization in the context of a $Z^\prime$ model is presented. The model is based on the $U(1)$ gauge group associated with the difference between muon- and tau-lepton number, $L_\mu - L_\tau$.}

\section{Introduction} \label{sec:intro}

Recently, the LHCb collaboration presented results from an angular analysis of the $B\to K^* \mu^+\mu^-$ decay based on 1/fb of data~\cite{Aaij:2013qta}. The results show a discrepancy in angular observables with respect to the Standard Model (SM) predictions. In particular, the observable $P_5^\prime$ (that corresponds to the observable $S_5$ in~\cite{Altmannshofer:2008dz}) shows a discrepancy with respect to the SM prediction of~\cite{Descotes-Genon:2013vna} with a local significance of 3.7$\sigma$ in the bin of di-muon invariant mass $4.3$~GeV$^2 < q^2 < 8.68$~GeV$^2$. In the $1$~GeV$^2 < q^2 < 6$~GeV$^2$ bin, that corresponds to the large recoil region under best theoretical control and that is used by default in many theory interpretations of the $B\to K^* \mu^+\mu^-$ data, the significance of the discrepancy is 2.5$\sigma$.
While unexpectedly large power corrections might be at least in part responsible for the observed discrepancy~\cite{Jager:2012uw}, it is interesting to interpret the experimental results in terms of new physics (NP) and to investigate what classes of NP models the current discrepancy favors.

In section~\ref{sec:independent}, based on~\cite{Altmannshofer:2013foa}, we discuss a model independent analysis of NP effects in the $B\to K^* \mu^+\mu^-$ decay and in all the relevant related $b \to s$ transitions. We identify which modifications of Wilson coefficients can lead to a consistent description of the available experimental data. In section~\ref{sec:Zprime}, based on~\cite{Altmannshofer:2014cfa}, we present an explicit $Z^\prime$ model capable of explaining the observed discrepancy.

\section{Model Independent Implications of the \texorpdfstring{$B \to K^* \mu^+\mu^-$}{B-->K*mu+mu-} Anomaly} \label{sec:independent}

The $B \to K^* \mu^+\mu^-$ decay and the related decays $B_s\to\mu^+\mu^-$, $B\to K\mu^+\mu^-$ and $B \to X_s\gamma$
are described by an effective Hamiltonian
\begin{equation}
\label{eq:Heff}
{\cal H}_{\rm eff} = - \frac{4\,G_F}{\sqrt{2}} V_{tb}V_{ts}^* \frac{e^2}{16\pi^2}
\sum_i (C_i O_i + C'_i O'_i) + {\rm h.c.} ~.
\end{equation}
It consists of flavor changing dimension 6 operators $O_i^{(\prime)}$ and the corresponding Wilson coefficients $C_i^{(\prime)}$.
We consider NP effects in the magnetic dipole operator $O_7$ and in the semileptonic operators $O_9$ and $O_{10}$ as well as in their chirality flipped counterparts $O_7^\prime$, $O_9^\prime$, and $O_{10}^\prime$
\begin{eqnarray}
O_7^{(\prime)} &=& \frac{m_b}{e}
(\bar{s} \sigma_{\mu \nu} P_{R(L)} b) F^{\mu \nu} ~, \\
O_9^{(\prime)} &=& 
(\bar{s} \gamma_{\mu} P_{L(R)} b)(\bar{\mu} \gamma^\mu \mu) ~, \label{eq:O9} \\
O_{10}^{(\prime)} &=&
(\bar{s} \gamma_{\mu} P_{L(R)} b)( \bar{\mu} \gamma^\mu \gamma_5 \mu) ~.
\label{eq:ops}
\end{eqnarray}
We do not consider NP effects in scalar, pseudo-scalar, or tensor operators, here. 
The distinct $q^2$ dependence of the discrepant $B\to K^*\mu^+\mu^-$ observable $S_5$ originates from the interference of contributions from the dipole operators and from the semileptonic operators. New Physics in either of them can bring $S_5$ in agreement with the data. 

However, finding a {\it consistent} explanation of the discrepancy in terms of NP is non-trivial. All the observables in $B\to K^*\mu^+\mu^-$ as well as the in the $B_s\to\mu^+\mu^-$, $B\to K\mu^+\mu^-$ and $B\to X_s\gamma$ decays, depend on the same Wilson coefficients. Therefore, a global analysis of model-independent constraints is required~\cite{DescotesGenon:2011yn}.
Here, we discuss results from our fit in~\cite{Altmannshofer:2013foa}, where details on the methodology and the used experimental data can be found. We mention that the latest $B \to K^*$ form factor results from the lattice~\cite{Horgan:2013hoa} as well as the latest LHCb results on the $B \to K \mu^+\mu^-$ and $B^+ \to K^{*+} \mu^+\mu^-$ branching ratios~\cite{Aaij:2014pli} are not yet included in this analysis.
Other recent model independent analyses can be found in~\cite{Descotes-Genon:2013wba}. 

\begin{figure} \centering
\includegraphics[width=0.31\textwidth]{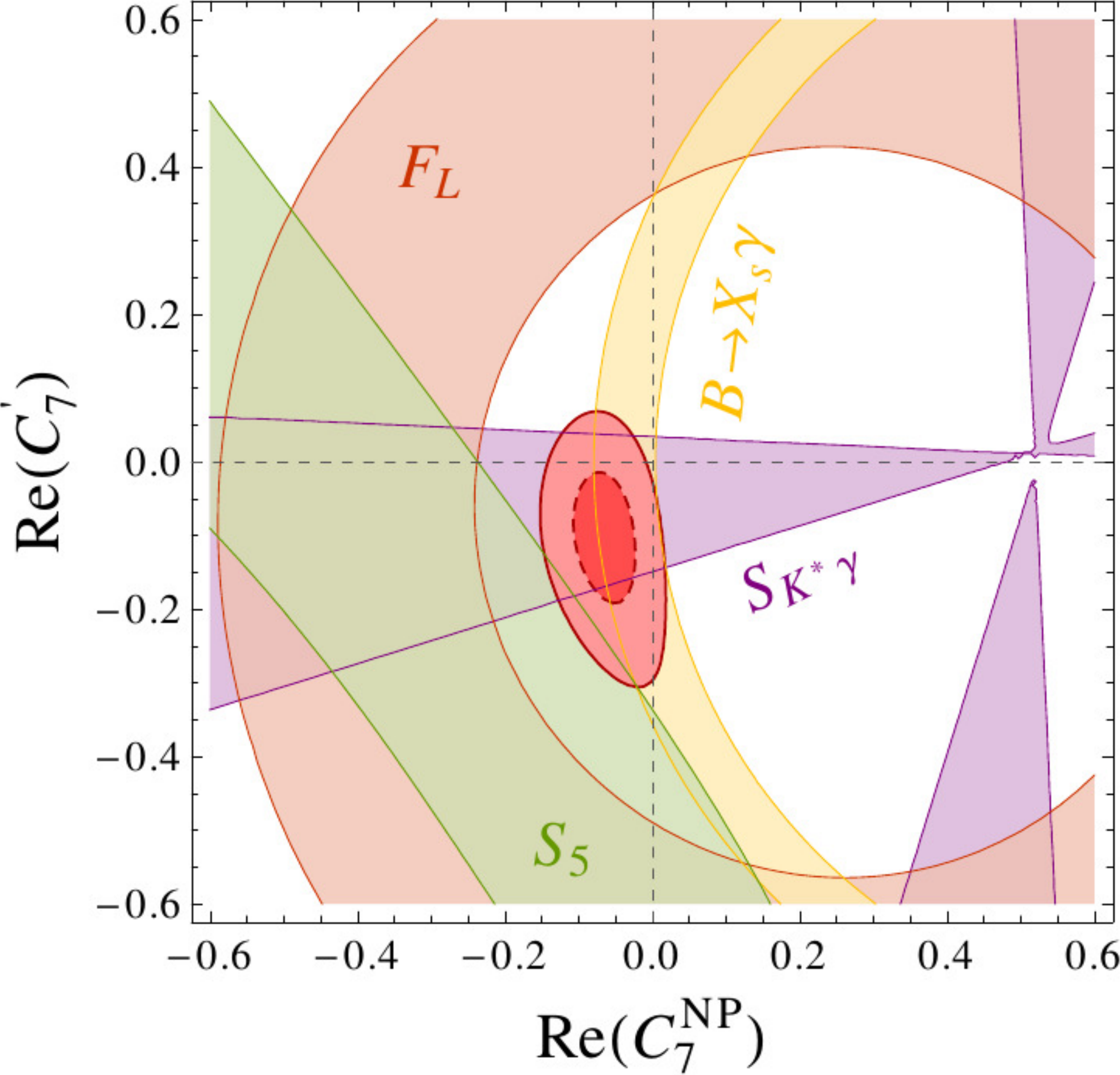} ~
\includegraphics[width=0.31\textwidth]{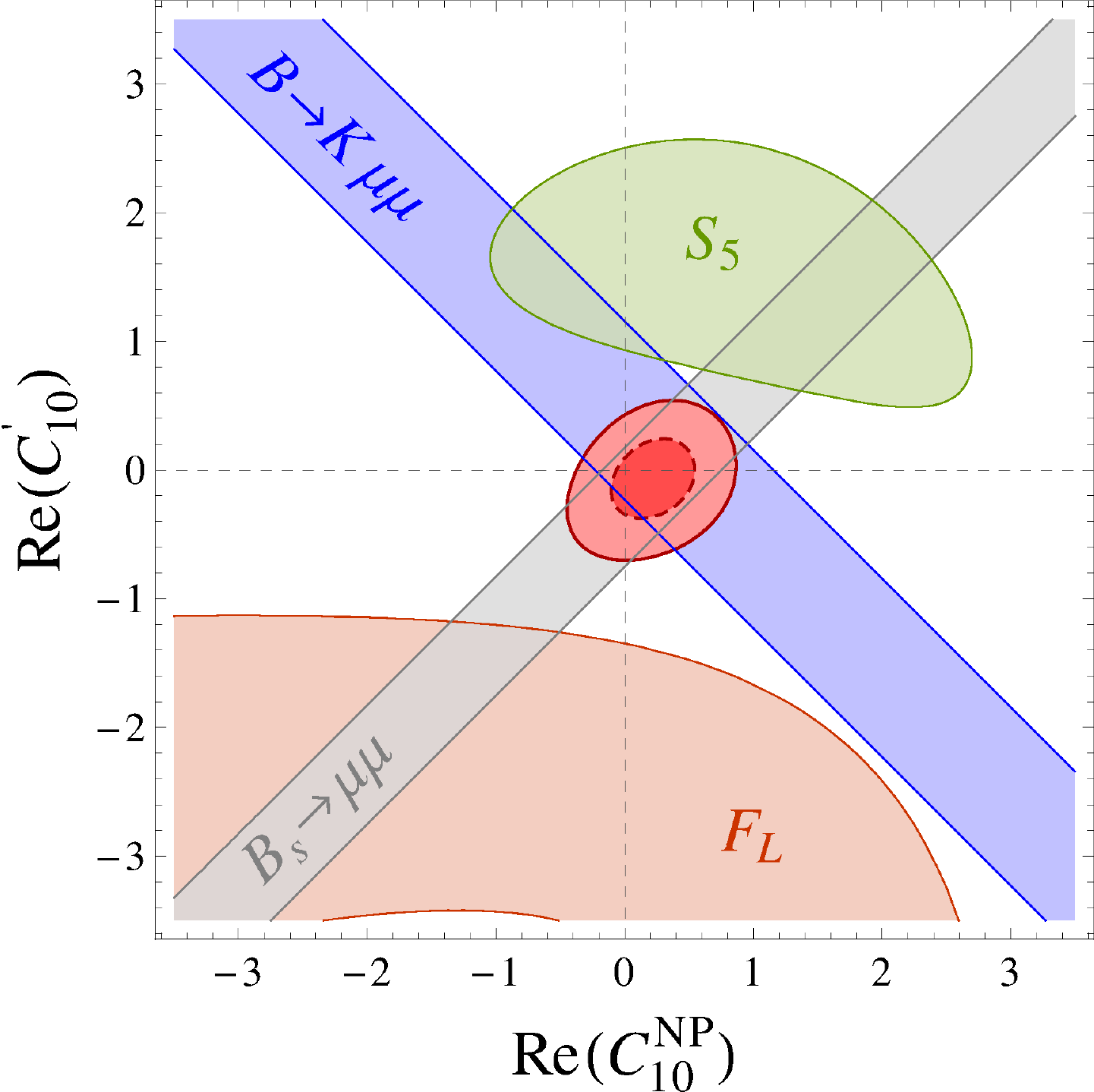} ~
\includegraphics[width=0.31\textwidth]{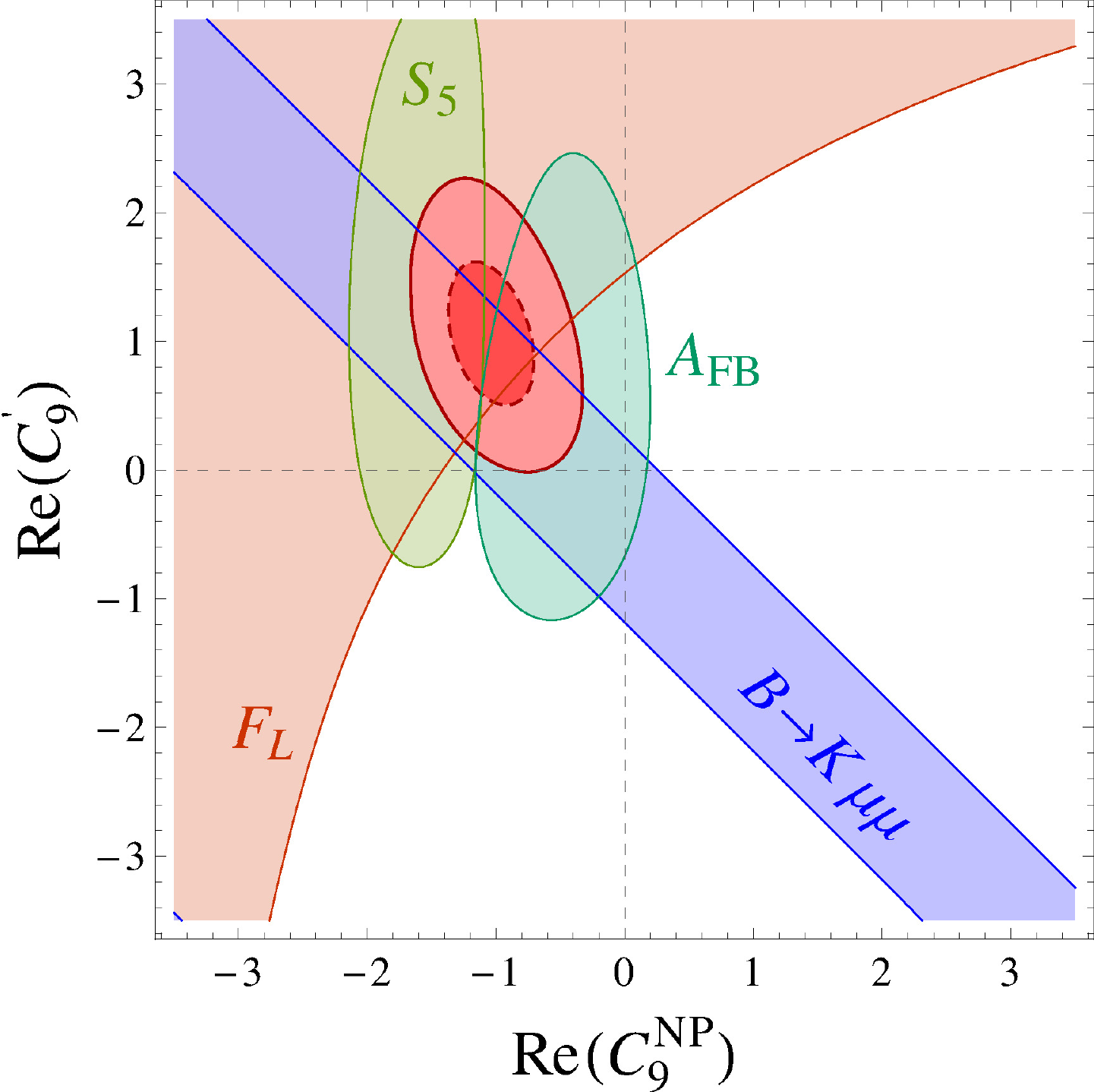}
\caption{Constraints in the $\text{Re}(C_7)-\text{Re}(C_7^\prime)$ plane (left), the $\text{Re}(C_{10}) - \text{Re}(C_{10}^\prime)$ plane (center), and the $\text{Re}(C_9) - \text{Re}(C_9^\prime)$ plane (right). Individual $\Delta \chi^2 = 1$ constraints are shown for BR$(B \to X_s \gamma)$ (yellow), $S_{K^*\gamma}$ (purple), $F_L$ (orange), $S_5$ (green), BR$(B \to K \mu^+\mu^-)$ (blue), BR$(B_s \to \mu^+\mu^-)$ (gray), and $A_\text{FB}$ (cyan). Combined $\Delta \chi^2 = 1,4$ contours are shown in red.}
\label{fig:fits}
\end{figure}

We discuss three scenarios: (i) real NP contributions to $C_7$ and $C_7^\prime$, (ii) real NP contributions to $C_{10}$ and $C_{10}^\prime$, and (iii) real NP contributions to $C_9$ and $C_9^\prime$.
We find that NP in $C_7$ and $C_7^\prime$ only cannot fully address the observed discrepancy. As shown in the left plot of Fig.~\ref{fig:fits}, the branching ratio of the $B \to X_s \gamma$ decay as well as the time dependent CP asymmetry in $B \to K^* \gamma$, $S_{K^* \gamma}$, strongly constrain NP in $C_7$ and $C_7^\prime$ and the tension in $S_5$ can only be improved slightly in scenario (i). Scenario (ii) is strongly constrained by the combination of experimental data on the $B \to K \mu^+ \mu^-$ and $B_s \to \mu^+\mu^-$ branching ratios as shown in the middle plot of Fig.~\ref{fig:fits}. The tension in $S_5$ cannot be explained by NP in $C_{10}$ and $C_{10}^\prime$. Finally, in scenario (iii), we find that a consistent explanation of the discrepancy is possible. As shown in the right plot of Fig.~\ref{fig:fits}, NP in the Wilson coefficient $C_9$ corresponding to $C_9^\text{NP} \sim -1.5$ (approximately $-35\%$ of the SM contribution) can account for the observed $S_5$. The constraint from BR$(B \to K \mu^+\mu^-)$ can be completely avoided by a NP contribution to $C_9^\prime$ of the same size but of opposite sign. An important constraint comes from the forward backward asymmetry in $B \to K^* \mu^+\mu^-$ (shown in cyan) that limits the allowed NP effects in $C_9$.
The best fit values for the Wilson coefficients read
\begin{equation}
 C_9^\text{NP} = -1.0 \pm 0.3 ~,~~ C_9^\prime = + 1.0 \pm 0.5 ~.
\end{equation}
Slightly better fits can be obtained by considering NP in all Wilson coefficients simultaneously and allowing also for CP violation. This however comes at the cost of a large number of free parameters.   

Focusing on the $C_9 - C_9^\prime$ scenario, we can translate the best fit values for the Wilson coefficients into a NP scale. Defining NP effects to the effective Hamiltonian by $ \Delta \mathcal H_\text{eff}= - \sum_i O_i/\Lambda_i^2$, the best fit values correspond to a scale
\begin{equation}
\label{eqn:effective_coeff}
|\Lambda_9| \simeq |\Lambda_9^\prime| \simeq 35~\text{TeV} ~.
\end{equation}
This is the scale of tree-level NP contributions with O(1) flavor changing $b \leftrightarrow s$ couplings and O(1) couplings to muons. If the NP effect arises at the 1-loop level, the scale is smaller by a factor of $4\pi$. Assuming minimal flavor violation, the scale is smaller by another factor of $\sqrt{1/|V_{ts}^*V_{tb}|} \simeq 5$.

\section{An Explicit \texorpdfstring{$Z^\prime$}{Z'} Model for the \texorpdfstring{$B \to K^* \mu^+\mu^-$}{B-->K*mu+mu-} Anomaly} \label{sec:Zprime}

The $B \to K^* \mu^+\mu^-$ anomaly is best explained by NP in the operators $O_9$ and $O_9^\prime$, that have vector couplings to muons, $(\bar\mu \gamma^\mu \mu)$, see~(\ref{eq:O9}). The presence of such operators, together with the absence of axial-vector and magnetic dipole operators, is intriguing as it cannot be realized in well-known extensions of the SM, like the minimal supersymmetric standard model (MSSM) or models with partial compositeness~\cite{Altmannshofer:2013foa}.
Most NP explanations of the anomaly make use of $Z^\prime$ gauge bosons. In particular, so-called $331$ models have been discussed extensively~\cite{Gauld:2013qba}.
Other promising candidates are $Z^\prime$ models based on the anomaly free $U(1)$ gauge group associated with the difference between muon- and tau-lepton number, $L_\mu - L_\tau$~\cite{He:1990pn}, which automatically leads to muonic vector-currents of the required type. Here we discuss the framework proposed in~\cite{Altmannshofer:2014cfa}. In order to give mass to the $Z^\prime$ boson, we introduce a scalar boson $\Phi$ that has $L_\mu - L_\tau$ charge and breaks $L_\mu - L_\tau$ spontaneously once it develops a vev $\langle \Phi \rangle = v_\Phi/\sqrt{2}$. This leads to a $Z^\prime$ mass $m_{Z^\prime} = g^\prime v_\Phi$, where $g^\prime$ is the $L_\mu - L_\tau$ gauge coupling.

\begin{figure}
\centering
\includegraphics[width=0.24\textwidth]{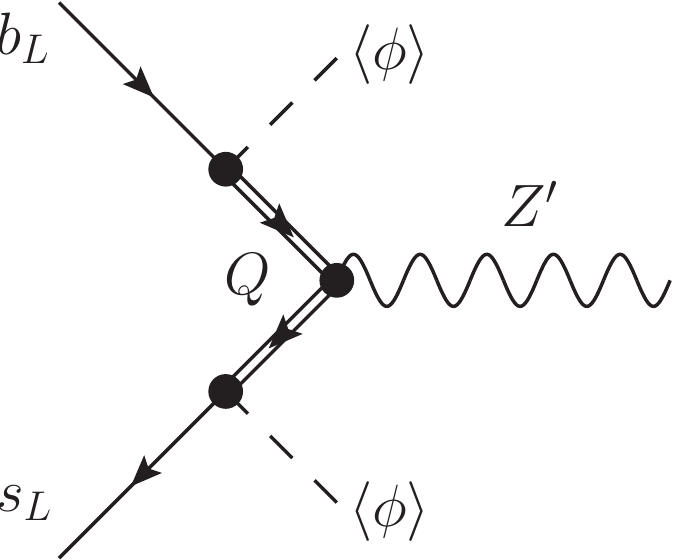} ~~~~~~~
\includegraphics[width=0.24\textwidth]{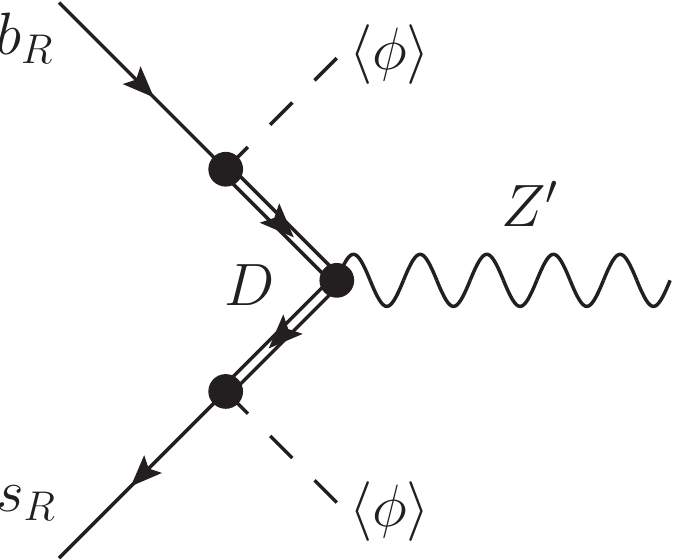} ~~~~~~~
\includegraphics[width=0.24\textwidth]{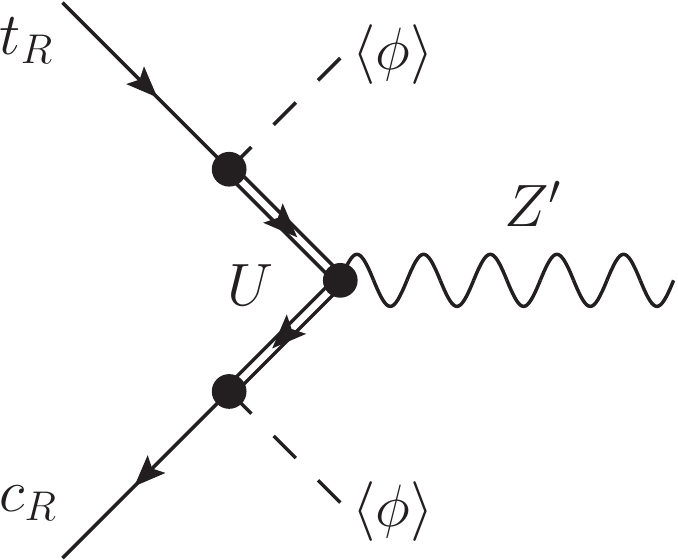}
\caption{Example diagrams that lead to flavor-changing couplings of the $L_\mu - L_\tau$ gauge boson to SM quarks.}
\label{fig:diagrams}
\end{figure}

In order to contribute to the $B \to K^* \mu^+\mu^-$ decay, the $Z^\prime$ has to couple to quarks as well. The required flavor changing couplings to quarks can be generated using an ``effective'' approach~\cite{Fox:2011qd}. We introduce one generation of heavy vector-like fermions $Q$, $U$, $D$, that are copies of the SM quarks, but carry $L_\mu - L_\tau$ charge such that they can couple to the SM quarks and the scalar $\Phi$. Once $\Phi$ develops a vev, the SM quarks and the vector-like quarks mix and effective flavor changing $Z^\prime$ quark couplings can be generated as shown in the diagrams of Fig.~\ref{fig:diagrams}.

Integrating out the $Z^\prime$ leads to the following contributions to $B\to K^*\mu^+\mu^-$
\begin{equation}
C_9 = \frac{Y_{Qb} Y_{Qs}^*}{2m_Q^2} ~,~~ C_9^\prime = -\frac{Y_{Db} Y_{Ds}^*}{2m_D^2} ~,
\end{equation}
where $Y_{Qb}$, for example, denotes the Yukawa coupling that mixes the vector-like quark $Q$ and the left-handed bottom quark $b_L$. Note that the Wilson coefficients $C_9$ and $C_9^\prime$ are independent of the $Z^\prime$ mass and the $U(1)^\prime$ gauge coupling\footnote{This is true as long as the $Z^\prime$ is sufficiently heavy compared to the decaying $B$ meson, such that the effective operator description in~(\ref{eq:Heff}) is valid.}.
In the following we assume a flavor structure for the mixing Yukawas $Y_{Qs} \sim Y_{Db} \sim 1$ and $Y_{Qs} \sim Y_{Ds} \sim \lambda^2$, where $\lambda \simeq 0.23$ is the Cabibbo angle. With this structure, an explanation of the $B \to K^* \mu^+\mu^-$ anomaly fixes the mass of the vector-like quarks to $m_Q \sim m_D \sim 5$~TeV. 

\begin{figure}
\centering
\includegraphics[width=0.48\textwidth]{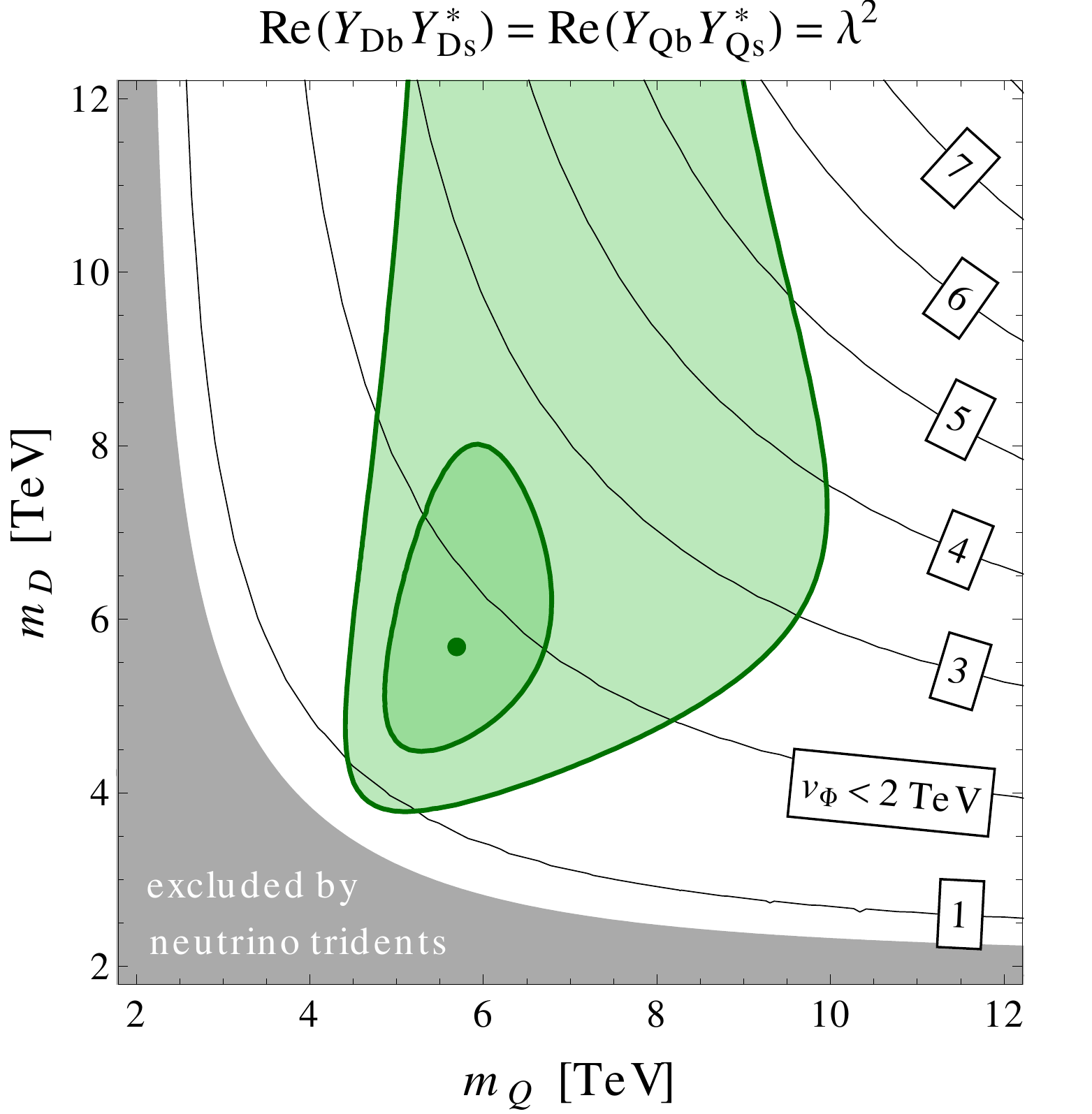} ~~~ 
\includegraphics[width=0.48\textwidth]{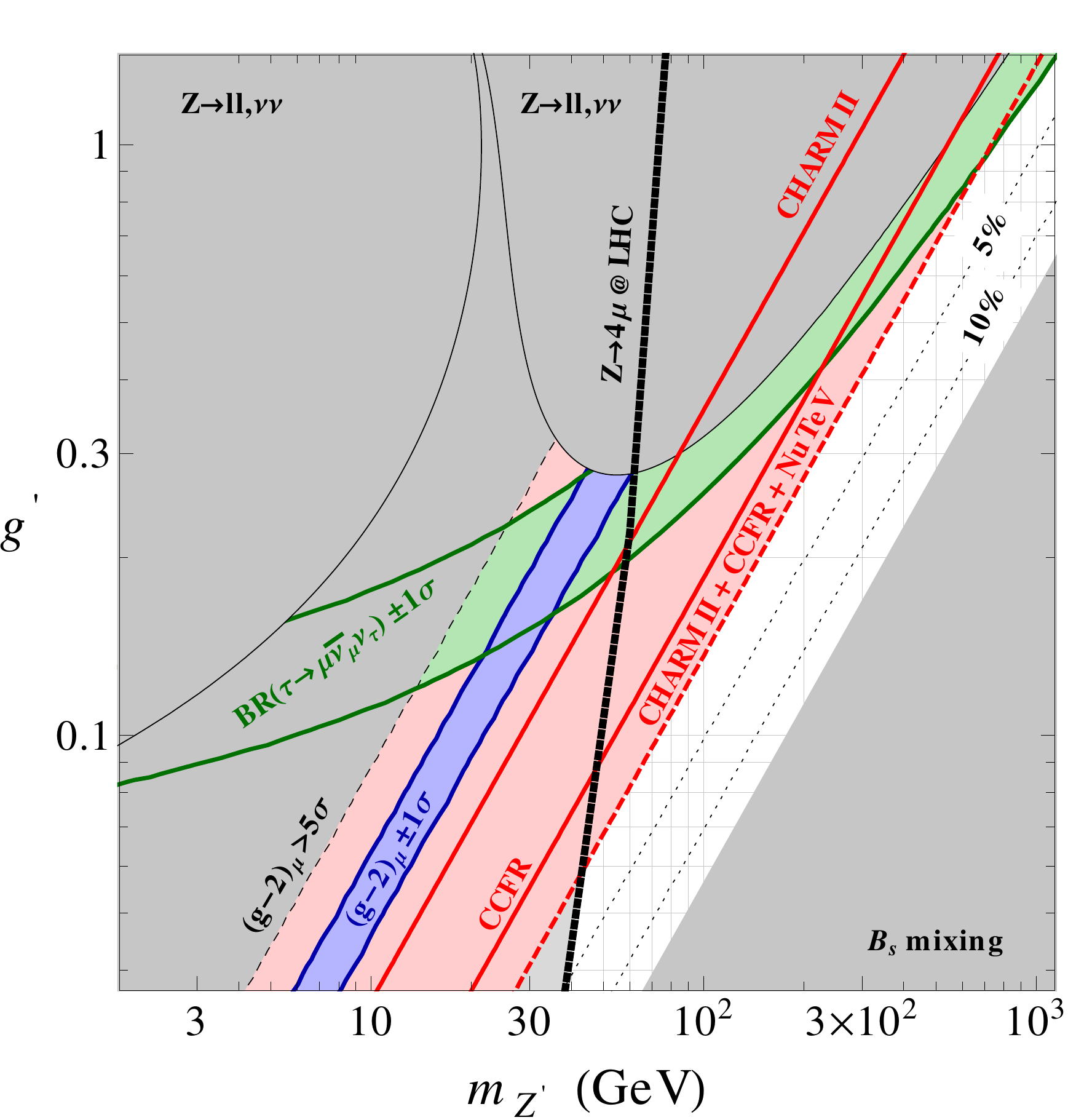}
\caption{Left: Constraints from $B_s$ mixing on the $U(1)^\prime$ breaking VEV, $v_\Phi$, in the plane of the vector-like quark masses $m_Q$ and $m_D$.  The region inside the green solid contours is preferred by the explanation of the $B \to K^* \mu^+\mu^-$ anomaly. The light gray region is excluded by experimental results on neutrino trident production. Right: Constraints on the $Z^\prime$ parameter space from various leptonic processes: the anomalous magnetic moment of the muon ``$(g-2)_\mu$'', leptonic tau decays ``BR$(\tau \to \mu \bar\nu_\mu \nu_\tau)$'', Z couplings to leptons and neutrinos ``$Z \to \ell\ell, \nu\nu$'', the measurement of $Z \to 4\mu$ at the LHC ``$Z \to 4\mu $@LHC'', and neutrino trident production ``CHARM-II + CCFR + NuTeV''. The allowed region is shown in white. The $B\to K^* \mu^+\mu^-$ anomaly can be accommodated everywhere to the left of the gray bottom-right triangle without being in conflict with $B_s$ mixing constraints. The dotted lines in the allowed region indicate the expected NP effects in $B_s$ mixing.}
\label{fig:Zprime}
\end{figure}

Integrating out the $Z^\prime$ also induces corrections to 4 fermion operators that mediate neutral meson mixing.  
Additional corrections can come from box diagrams involving the scalar $\Phi$ and the vector-like quarks. However, with the assumed flavor structure of the mixing Yukawas, the dominant contribution to meson mixing arises from tree-level exchange of the $Z^\prime$. The $Z^\prime$ contributions are proportional to $v_\Phi^2$. Allowing for at most 15\% NP in $B_s$ mixing, and assuming that the $Z^\prime$ explains the $B \to K^* \mu^+\mu^-$ anomaly, we find the {\it upper bound} $v_\Phi \lesssim 1.8$~TeV which corresponds to a $Z^\prime$ mass of $m_{Z^\prime} \lesssim g^\prime \cdot 1.8$~TeV. The upper bound on $v_\Phi$ in the $m_Q$ - $m_D$ plane is shown in the left plot of Fig.~\ref{fig:Zprime}.

Bounds from neutral Kaon and D-meson mixing restrict the couplings of the $Z^\prime$ to first generation quarks to be very small. Consequently, direct production of the $Z^\prime$ at hadron colliders is strongly suppressed and $Z^\prime$ searches at Tevatron and the LHC do not lead to relevant constraints.

However, the leptonic phenomenology of the $L_\mu - L_\tau$ gauge symmetry is rich and allows to probe large parts of parameter space of the considered model. Important probes include the $g-2$ of the muon, the leptonic tau decays $\tau \to \mu \nu_\tau \bar\nu_\mu$ and $\tau \to e \nu_\tau \bar\nu_e$, the couplings of the SM $Z$ boson to taus, muons and neutrinos, as well as the branching ratio of the SM $Z$ boson to four muons.
A particularly powerful constraint on the $L_\mu - L_\tau$ gauge boson arises form neutrino trident production, {\it i.e.} the production of a muon anti-muon pair in the scattering of muon neutrinos on a target nucleus. Integrating out the $Z^\prime$, which is a valid approximation for $Z^\prime$ masses of $m_{Z^\prime} \gtrsim 10$~GeV, we find for the trident cross section
\begin{equation}
\frac{\sigma}{\sigma_\text{SM}} \simeq \frac{1 + \left( 1 + 4 s_W^2 + 2v^2/v_\Phi^2 \right)^2 }{1 + \left( 1 + 4 s_W^2 \right)^2} ~.
\end{equation} 
Using the available experimental data on neutrino tridents~\cite{Geiregat:1990gz} we obtain a {\it lower bound} on the $L_\mu - L_\tau$ breaking vev $v_\Phi \gtrsim 750$~GeV. 
A summary of all the leptonic constraints of the $Z^\prime$ is shown in the right plot of Fig.~\ref{fig:Zprime}.
It will be very interesting to understand to which extent future neutrino experiments can probe the region of parameter space that is currently still unconstrained.

\section{Conclusions}

Rare $B$ decays play a central role among the indirect probes of new physics.
Interestingly enough, recent LHCb results on the $B\to K^* \mu^+\mu^-$ decay show a discrepancy with SM predictions.
A consistent explanation of this discrepancy in terms of new physics is possible as confirmed by various model-independent analyses~\cite{Altmannshofer:2013foa,Descotes-Genon:2013wba}.
The required new physics operators are readily accommodated in models that contain massive $Z^\prime$ gauge bosons with flavor changing $b \to s$ couplings as well as vector couplings to muons.
A promising candidate is a $Z^\prime$ boson that is associated to gauging the difference between muon- and tau-lepton number $L_\mu - L_\tau$~\cite{Altmannshofer:2014cfa}.  
In contrast to most of the $Z^\prime$ models that have been discussed in the literature in connection with the $B \to K^* \mu^+\mu^-$ discrepancy~\cite{Gauld:2013qba}, which envision very heavy $Z^\prime$'s, above $\sim 3$~TeV, the proposed $L_\mu - L_\tau$ gauge boson can be much lighter, even well below the electro weak scale. 
A very distinct feature of the discussed model is the pattern of lepton-flavor universality violation in $B$ decays. While the electron modes based on $b \to s e^+e^-$ are unaffected by the $Z^\prime$, observables in the muonic and tauonic modes based on the $b \to s \mu^+\mu^-$ and $b \to s \tau^+\tau^-$ transitions are modified by the same amount but with opposite signs. 

If the observed discrepancy in the $B\to K^* \mu^+\mu^-$ decay will be confirmed by an experimental analysis of the full LHCb data set, future precision measurements of the inclusive $B \to X_s \mu^+\mu^-$ decay and the neutrino modes based on $b \to s \nu\bar\nu$, as well as lepton flavor universality tests with $b \to s e^+e^-$ and $b \to s \tau^+\tau^-$ transitions will be invaluable in identifying a possible underlying new physics origin.

\section*{Acknowledgments}

I would like to thank the organizers for the kind invitation to the Moriond Electro Weak 2014 conference. I thank Stefania Gori, Maxim Pospelov, David Straub, and Itay Yavin for the fruitful collaborations these proceedings are based on. I also thank Stefania Gori for a careful reading of the manuscript. The research of WA was supported by the John Templeton Foundation. Research at Perimeter Institute is supported by the Government of Canada through Industry Canada and by the Province of Ontario through the Ministry of Economic Development \& Innovation.

\section*{References}

\end{document}